\documentclass[pra,aps,twocolumn,10pt]{revtex4}
\usepackage{amsmath,mathrsfs}
\usepackage{graphicx,epsfig}
\usepackage{latexsym,amssymb}
\usepackage{epsf,longtable,array}
\usepackage{ifpdf,natbib,lineno}
\newcommand{\sigmaLV}{\sigma_{\mathrm{LV}}}
\newcommand{\sigmaSL}{\sigma_{\mathrm{SL}}}
\newcommand{\sigmaSV}{\sigma_{\mathrm{SV}}}
\newcommand{\thetaY}{\theta_{\mathrm{Y}}}
\newcommand{\thetaW}{\theta^*_{\mathrm{W}}}
\newcommand{\thetaC}{\theta^*_{\mathrm{C}}}
\newcommand{\equref}[1]{Eq.~(\ref{#1})}

\newcommand{\figref}[1]{Fig.~\ref{#1}}
\newcommand{\Figref}[1]{Figure~\ref{#1}}
\newcommand{\figsref}[2]{Figs.~\ref{#1}~and~\ref{#2}}

\newcommand{\phiLeft}{\phi_{\mathrm{Left}}}
\newcommand{\phiRight}{\phi_{\mathrm{Right}}}
\newcommand{\psib}{\psi_{\mathrm{b}}}
\newcommand{\psis}{\psi_{\mathrm{s}}}
\newcommand{\rhos}{\rho_{\mathrm{s}}}

\newcommand{\rhoc}{\rho_{\mathrm{c}}}
\newcommand{\rhoL}{\rho_{\mathrm{L}}}
\newcommand{\rhoV}{\rho_{\mathrm{V}}}
\newcommand{\pc}{p_{\mathrm{c}}}

\newcommand{\Tc}{T_{\mathrm{c}}}
\newcommand{\pL}{p_{\mathrm{Laplace}}}

\newcommand{\delete}[1]{}
\begin{document}


\title{{\bf Roughness gradient induced spontaneous motion of droplets on hydrophobic surfaces: A lattice Boltzmann study}}

\author{Nasrollah Moradi$^{1}$ \footnote[1]{nasrollah.moradi@rub.de}, Fathollah Varnik$^{1,2}$ and Ingo Steinbach$^{1}$ }
\affiliation{$^{1}$\textit{ICAMS}, Ruhr-Universit\"at Bochum, Stiepeler Strasse 129, 44801 Bochum, Germany}
\affiliation{$^{2}$Max-Planck Institut f\"ur Eisenforschung, Max-Planck Str.~1, 40237 D\"usseldorf, Germany}

\date{\today}


\begin{abstract}
\par The effect of a step wise change in the pillar density on the dynamics of droplets is investigated via three-dimensional lattice Boltzmann simulations. For the same pillar density gradient but different pillar arrangements, both motion over the gradient zone as well as complete arrest are observed. In the moving case, the droplet velocity scales approximately linearly with the texture gradient. A simple model is provided reproducing the observed linear behavior. The model also predicts a linear dependence of droplet velocity on surface tension. This prediction is clearly confirmed via our computer simulations for a wide range of surface tensions.
\end{abstract}\maketitle

\section{Introduction}\label{action}
\par Due to its occurrence in a wide range of natural phenomena and its fundamental importance for surface engineering \cite{Quere, Dorrer}, the behavior of liquid drops on solid surfaces is an active field of research \cite{QuereAnnu}. Individual droplets can serve as ideal chemical reactors \cite{Ajdari}, carriers of information \cite{Prakash} or in ink-jet printers as well as surface preparation prior to painting or coating.

Although there exist a few solids that are molecularly flat (e.g.\ mica), most of solids are rough on the micro scale \cite{Rabbe} so that the well known Young's law, $\cos\thetaY=(\sigmaSV-\sigmaSL)/\sigmaLV$ \cite{Young}, \delete{describing the behavior of a droplet on a perfectly flat and chemically homogeneous surface} must be modified in order to take account of surface roughness. Here, $\thetaY$ is the contact angle on a flat substrate and $\sigmaLV$, $\sigmaSL$ and $\sigmaSV$ are the  liquid-vapor, solid-liquid and solid-vapor specific surface free energies, respectively. The simplest and most popular modification of the Young's law for rough surfaces dates back to the  works of Wenzel \cite{Wenzel} and Cassie and Baxter \cite{Cassie}, where the effect of roughness on wetting is assumed to be a mere change of the average surface areas involved in the problem.

Assuming that the liquid completely penetrates into the roughness grooves (collapsed state), Wenzel obtained $\cos\thetaW=r\cos\thetaY$ for the apparent contact angle $\thetaW$ (the roughness factor $r$ is the real solid area within a square of unit length). Cassie and Baxter, on the other hand, considered the case of a droplet pending on the top of roughness tips  (suspended state) and obtained
\begin{equation}
\cos\thetaC=\phi\cos\thetaY-(1-\phi),
\label{eq:Cassie}
\end{equation}
where the roughness density $\phi$ gives the fraction of the
droplet's base area, which is in contact with the solid. It is
important to realize that both the Wenzel and the Cassie-Baxter
equations do not explicitly take account of three phase contact
line structure. This shortcoming may, however, be neglected as
long as the contact area reflects the structure and energetics of
the three phase contact line \cite{Gao2007a}.

The suspended state is often separated from the Wenzel state by a finite free energy barrier, which depends both on the  droplet size and roughness characteristics \cite{Reyssat1, Jopp, Markus}. On a rough hydrophobic substrate, the apparent contact angle of a droplet in the suspended state is typically higher than in the collapsed state \cite{ Dorrer, Li, Oner}.
Furthermore, the contact angle hysteresis significantly increases when a suspended droplet undergoes a transition to the Wenzel state  \cite{Quere,QuereAnnu}. Indicative of stronger pinning \cite{Joanny} of the three phase contact line, this feature reflects itself in a sticky behavior of liquid drops in the collapsed state \cite{Dorrer, Lafuma} as compared to their high mobility in the suspended state.

While droplet behavior on homogeneous roughness has widely been investigated in the literature, only few works exist dealing with the case of \emph{inhomogeneous} topography \cite{Zhu, Yang, Shastry, Reyssat, Fang}. Indeed, experimental observation of a roughness gradient induced spontaneous motion is not an easy task \cite{Reyssat}. The authors of \cite{Shastry,Reyssat}, for example, resort to shaking vertically the substrate in order to overcome pinning forces. 

Recently, a two-dimensional theoretical model is proposed aiming at a study of the present topic \cite{Fang}. To the best of our knowledge, however, computer simulations of the problem are lacking so far. The present work is aimed at filling this gap. We provide first direct numerical evidence for spontaneous droplet motion actuated by a gradient of pillar density. Furthermore, we investigate the influence of specific distribution/arrangement of roughness elements (pillars in our case) on the behavior of the droplet. An important observation is that, depending on the specific arrangement of the pillars, both complete arrest and motion over the entire gradient zone can be observed for the same gradient of pillar density. This underlines the importance of the topography design for achieving high mobility drops.

For the case of mobile drops, we provide a simple model for the dependence of drop velocity on pillar density difference $\Delta \phi = \phiRight-\phiLeft$. The model accounts for the observed linear dependence and predicts further that the velocity should scale linearly with the liquid-vapor surface tension. This prediction is also in line with our computer simulations for the studied range of parameters.


\section{Numerical Model}
We employ a free energy based two phase lattice Boltzmann (LB) method, first proposed by Swift \cite{Swift}. After establishing the Galilean invariance \cite{Holdych}, it was developed further \cite{Briant,Dupuis} in order to take the wetting effect of the solid substrate into account. Since then, the approach has been used to study e.g.\ stability and dynamics of droplets on topographically patterned hydrophobic substrates \cite{Kusumaatmaja,Dupuis,Yeomans}, effect of chemical surface patterning on droplet dynamics \cite{Kusumaatmaja2,Yeomans} as well as chemical gradient induced separation of emulsions \cite{VarnikF}. A detailed description of the method can be found in \cite{Briant,Dupuis}. For the sake of completeness, however, we present a short overview of the method. Equilibrium properties of the present model can  be obtained from a free energy functional
\begin{equation}
\Psi = \int_{V} \left(\psib(\rho({\bf r}))+\frac{\kappa}{2} (\partial_{\alpha}\rho({\bf r}))^2 \right) d{\bf r}^3 + \int_{S} \psis ds.
\label{eq_free_energy_model}
\end{equation}
In \equref{eq_free_energy_model}, $\psib$ is the bulk free energy density of the fluid, $V$ denotes the system volume, $S$ is the substrate surface area and  $\rho({\bf r})$ the fluid density at point ${\bf r}$. Considering a simple van der Waals model  \cite{Briant}, the bulk free energy per unit volume,  $\psib (\rho)$, can be given as $\psib (\rho) = \pc (\nu_\rho+1)^2 (\nu_\rho^2-2\nu_\rho+3-2\beta\nu_T)$ in which $\nu_\rho = (\rho-\rhoc)/\rhoc$  and  $\nu_T = (\Tc-T)/\Tc$ are  the reduced density and reduced temperature, respectively.  The critical density, pressure, and temperature are set to $\rhoc=7/2$, $\pc=1/8$, and $\Tc=4/7$, respectively.  Below $\Tc$, the model describes liquid-vapor coexistence with related  equilibrium densities  $\rho_{\mathrm{L,V}}=\rhoc(1\pm\sqrt{\beta\nu_{T}})$.

The parameter $\beta$ is related to the interface thickness $\xi$ and  the surface tension $\sigma$ via $\xi=\sqrt{\kappa\rhoc^{2}/(4\beta\nu_T\rhoc)}$ and $\sigma=4/3\sqrt{2\kappa pc}(\beta\nu_T)^{3/2}\rhoc$. When combined with an appropriate variation of $\kappa$, it allows to vary the surface tension and interface width independently. Using Cahn-Hilliard approach, the surface free energy per unit area, $\psis$, is approximated as  $-\phi_{1}\rhos $, where $\rhos$ is the density of  fluid on the solid substrate and $\phi_{1}$ is a constant which can be used to tune the contact angle.  Minimizing the free energy functional $\Psi $, \equref{eq_free_energy_model}, subject to the condition $\psis=-\phi_{1}\rhos $ leads to an equilibrium boundary condition for the spatial derivative of fluid density in the direction normal to the substrate, $\partial_{\perp}\rho=-\phi_{1}/\kappa$. The parameter $\phi_{1}$ is related to the Young contact angle via
\begin{eqnarray}
\phi_{1}=2\beta\tau_{T}\sqrt{2\pc\kappa} \mathrm{sign} (\frac{\pi}{2}-\theta)\sqrt{\cos\frac{\alpha}{3}-(1-\cos\frac{\alpha}{3})},
\label{eq02}
\end{eqnarray}
where  $\alpha=\cos^{-1}(\sin^{2}\thetaY)$, and $\thetaY$ is the equilibrium Young contact angle and the function ``sign'' determines the sign of its argument.  All the quantities in this paper are given in  dimensionless lattice Boltzmann units.
The LB relaxation time is set to $\tau=0.8$ and the temperature is fixed to $T=0.4$. For a typical choice of $\beta=0.1$, for example, this leads to the equilibrium liquid and vapor densities $\rhoL \approx$ 4.1 and $\rhoV \approx$ 2.9. Depending on the case of interest, $\kappa$ lies in the range $[0.002, 0.008]$ and the size of the simulation box is varied with values around $L_{x}\times L_{y}\times L_{z} =$ 125 $\times$ 90  $\times$ 90 lattice nodes for spherical, and  $L_{x}\times L_{y}\times L_{z} =$ 125 $\times$ 25  $\times$ 90 for cylindrical droplets. Periodic boundary conditions are applied in the $x$ and $y$-directions.

\section{Discussion and Results}
As mentioned above, in the experimental reports which have considered the motion of a suspended droplet on surfaces with a gradient of texture, the behavior of droplets is not unique \cite{Yang,Zhu,Shastry,Reyssat}. This indicates that the roughness factor as well as the roughness density are not sufficient for a full characterization of a rough surface. Although \equref{eq:Cassie} predicts a decrease of the effective contact angle upon an increase of roughness density and hence a driving force along the gradient of $\phi$, the contact angle hysteresis~\cite{Kusumaatmaja3} may be strong enough in order to prevent a spontaneous droplet motion \cite{Reyssat}.

The present work underlines this aspect by explicitly showing that
the behavior of a droplet on substrates patterned by pillar
microstructure with the same pillar density gradient, but
different pillar geometries (e.g.~rectangular posts with different
pillar width and spacing) can be qualitatively different. While in
the one case the droplet spontaneously moves due to the roughness
gradient induced driving force, it may become arrested if an
unfavorable geometry is chosen.

\begin{figure}
 \includegraphics[width=4.25 cm]{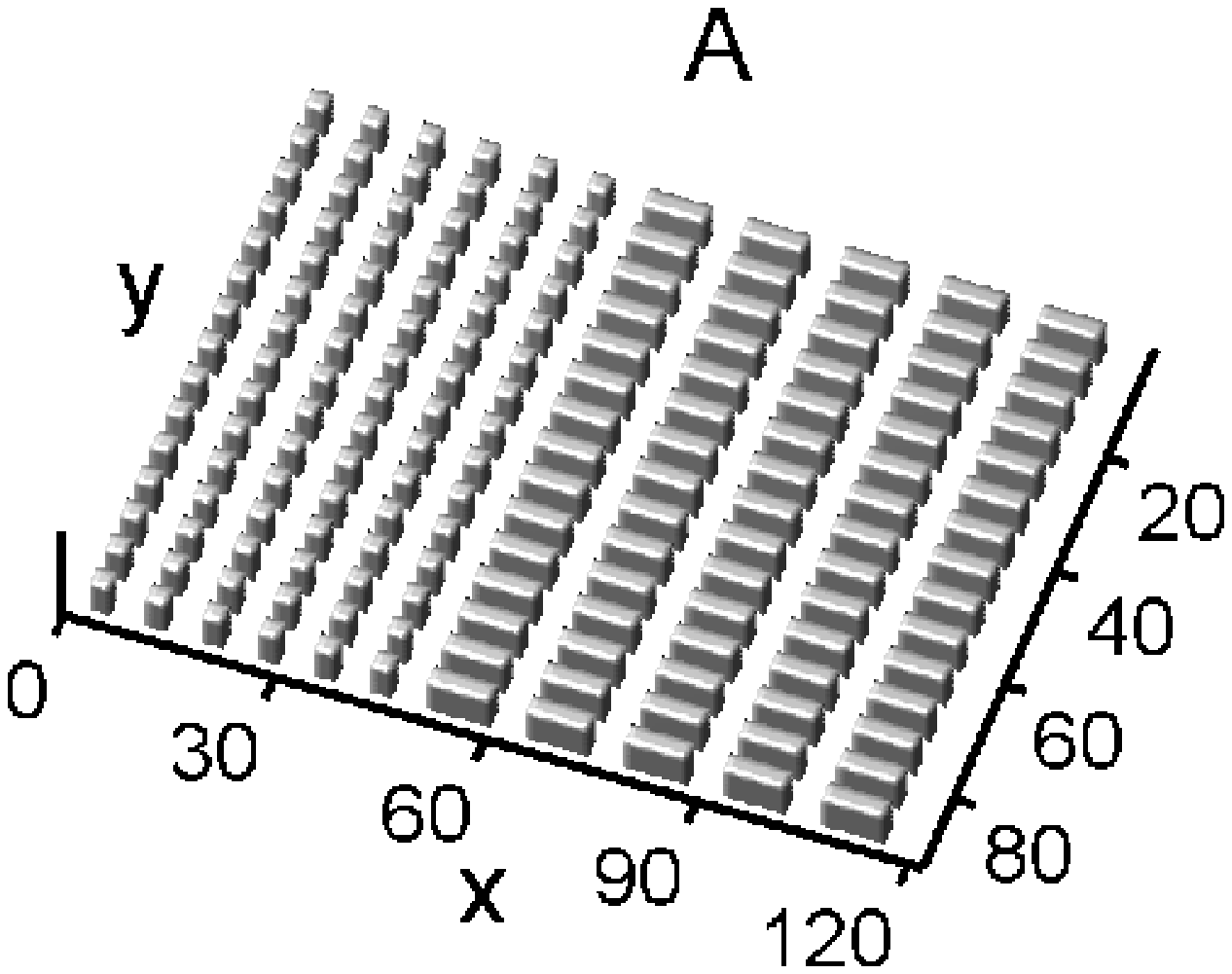}
 \includegraphics[width=4.25 cm]{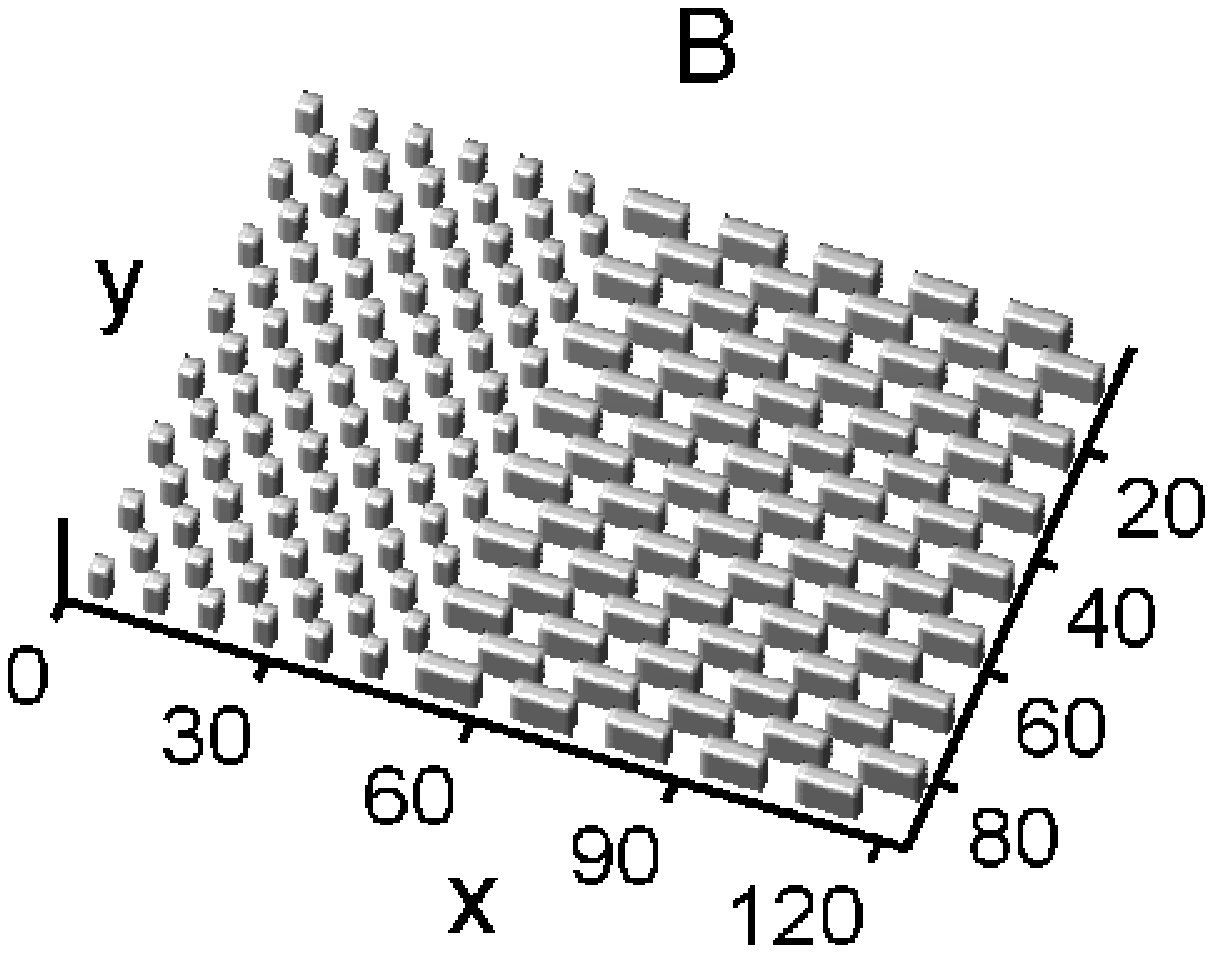}
\caption{Top view of two step gradient substrates. In the left panel (referred to as case A) , the pillar density to the left side ($x<50$) is $\phiLeft=0.187$ (square posts of length $a=b=3$) while it is set to $\phiRight=0.321$ on the right side ($x>50$, rectangular posts of length $a=9$ and width $b=3$). The spacing distance of the pillars in the $x$-direction is $d_x=5$ and in the $y$-direction is $d_y=3$ overall on the substrate. The height of the posts is $c=6$. The right panel (case B) is obtained from A by shifting the posts on each second raw horizontally by an amount of $(a+d_x)/2$ with $d_x=5$, $a=3$ for $x<50$ and $a=9$ for $x>50$. All lengths are given in  LB units.}
\label{fig:substrates}
\end{figure}

A simple way to study the effect of a gradient texture is to introduce an abrupt (stepwise) change in the roughness (pillar) density along a given spatial direction. Adopting this choice, we design a substrate divided into two regions, each with a constant pillar density. In order to underline the crucial role of pillar arrangement on the behavior of droplet, we consider two different cases of the same pillar density gradient as shown in \figref{fig:substrates}.


\begin{figure}
\includegraphics[width=4.25 cm]{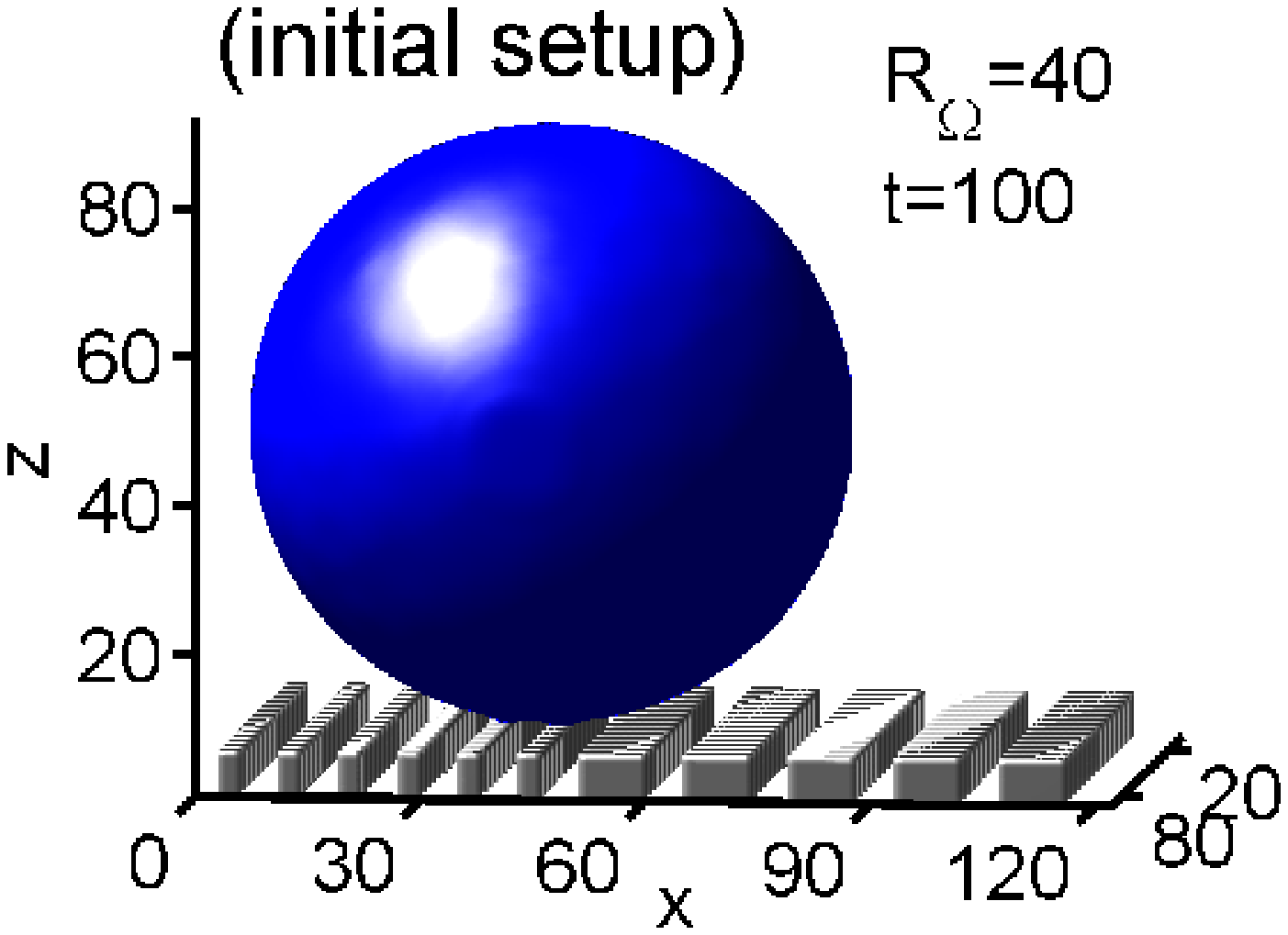}
\includegraphics[width=4.25 cm]{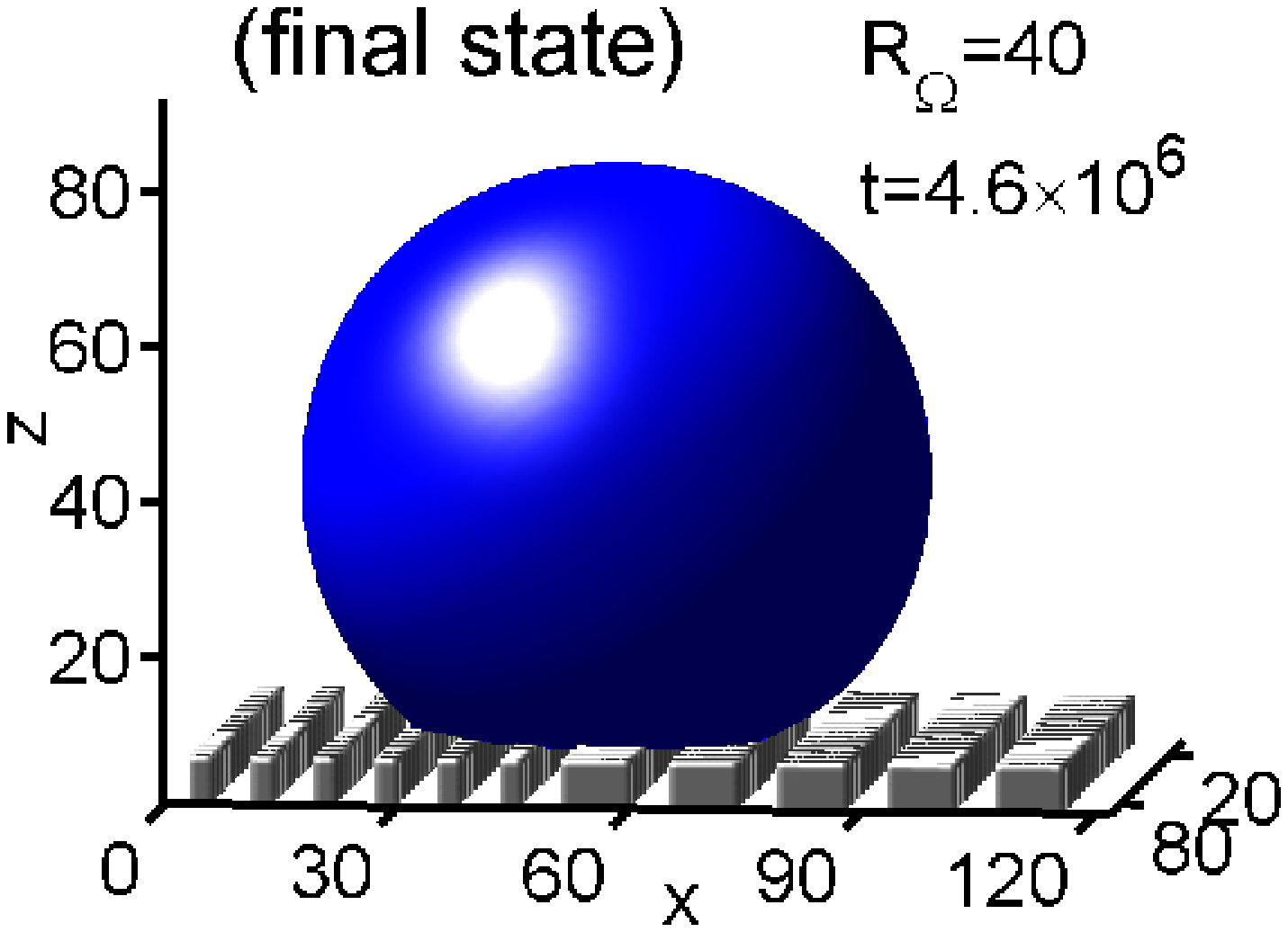}
\includegraphics[width=4.25 cm]{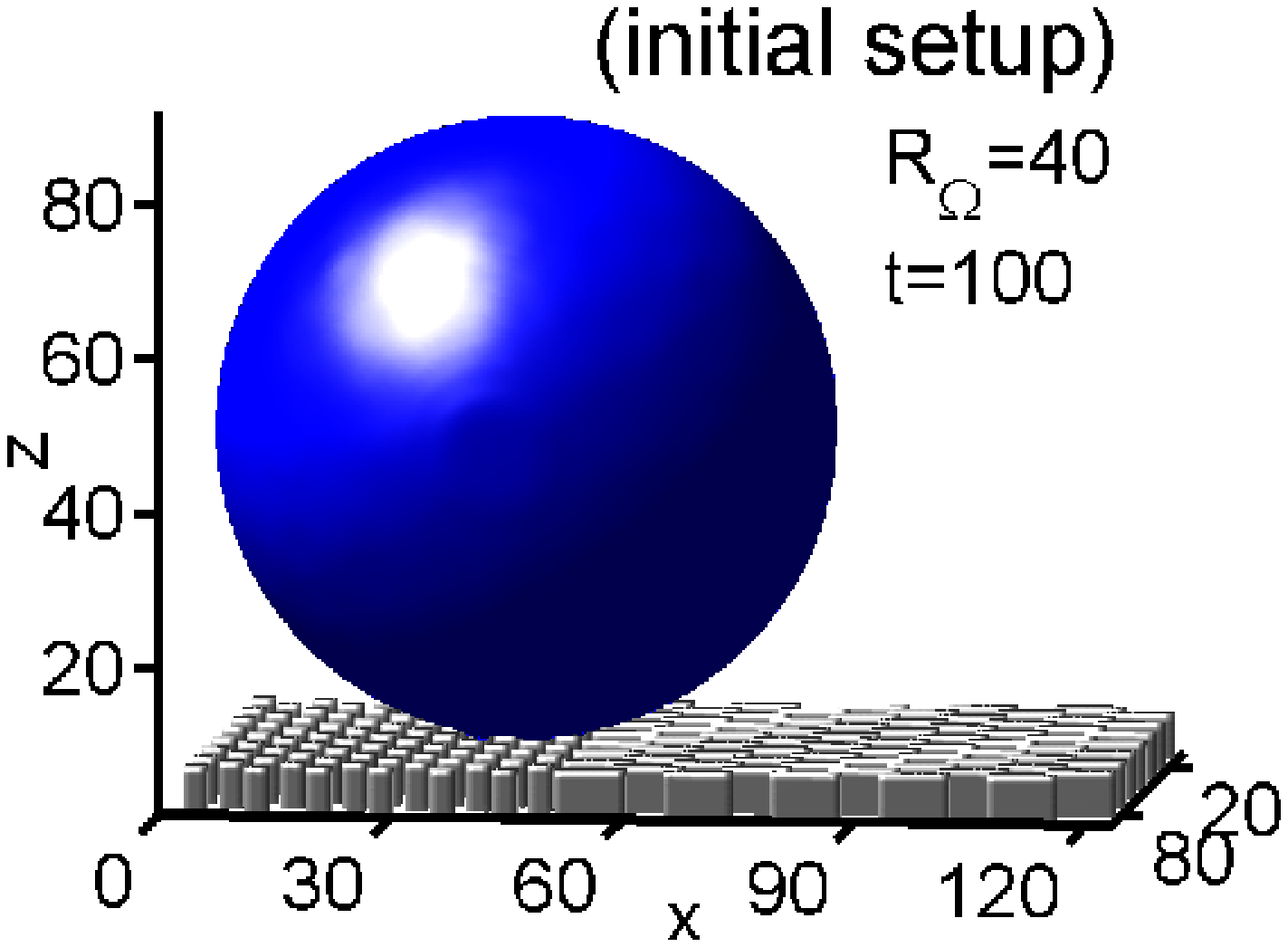}
\includegraphics[width=4.25 cm]{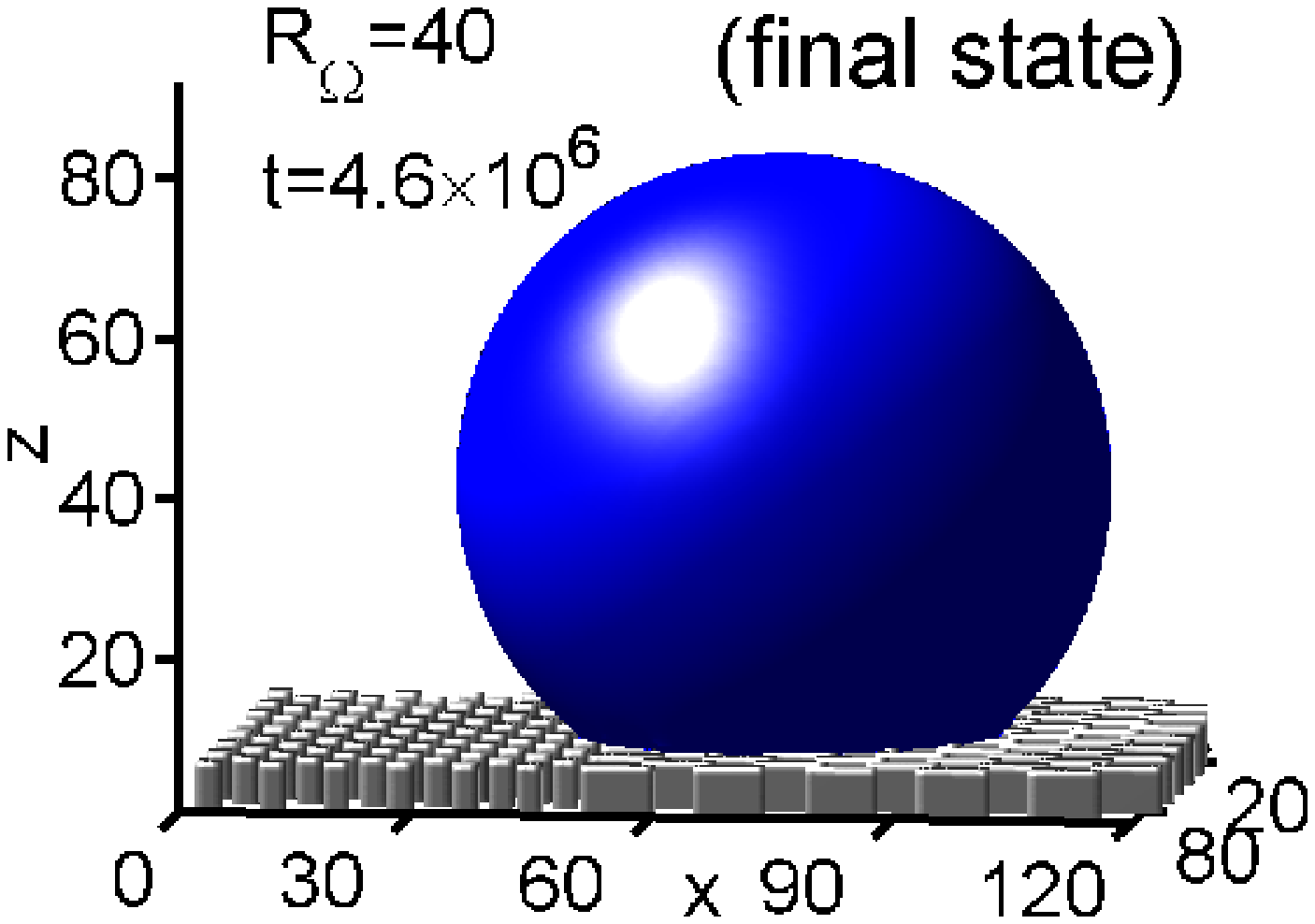}
\caption{Initial setup and final states of a spherical droplet on substrates with an abrupt (step-wise) change of pillar density. The cases of substrates A and B (see \figref{fig:substrates}) are compared.}
\label{fig:sph_drop_snapshots}
 \end{figure}

Using the two substrates shown in \figref{fig:substrates}, we performed a series of lattice Boltzmann simulations placing at time $t=0$ a spherical liquid droplet close to the top of the border line separating the two regions of different pillar density. A close look at the left panels in \figref{fig:sph_drop_snapshots} reveals that, both in the case of substrates A and B, the presence of a roughness gradient leads to an asymmetric spreading of droplet. However, despite this similarity of the dynamics at the early stages of spreading, the long time behavior of the droplet strongly depends on the specific arrangement of pillars. In particular, in the case of substrate A, the droplet motion is stopped on the gradient zone, while in case of substrate B it completely reaches the more favorable region of higher $\phi$.

\begin{figure}
\includegraphics[width=4.25 cm]{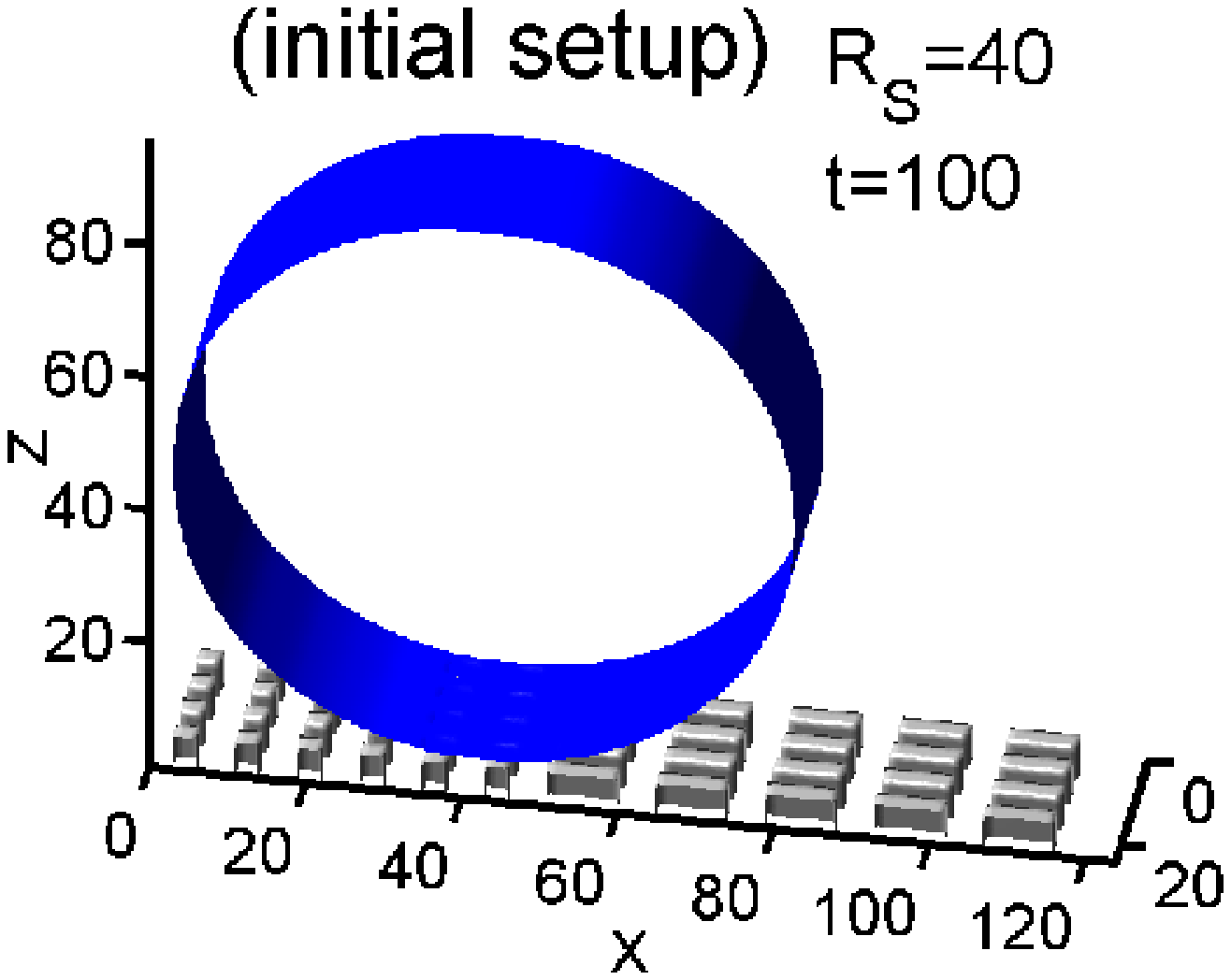}
\includegraphics[width=4.25 cm]{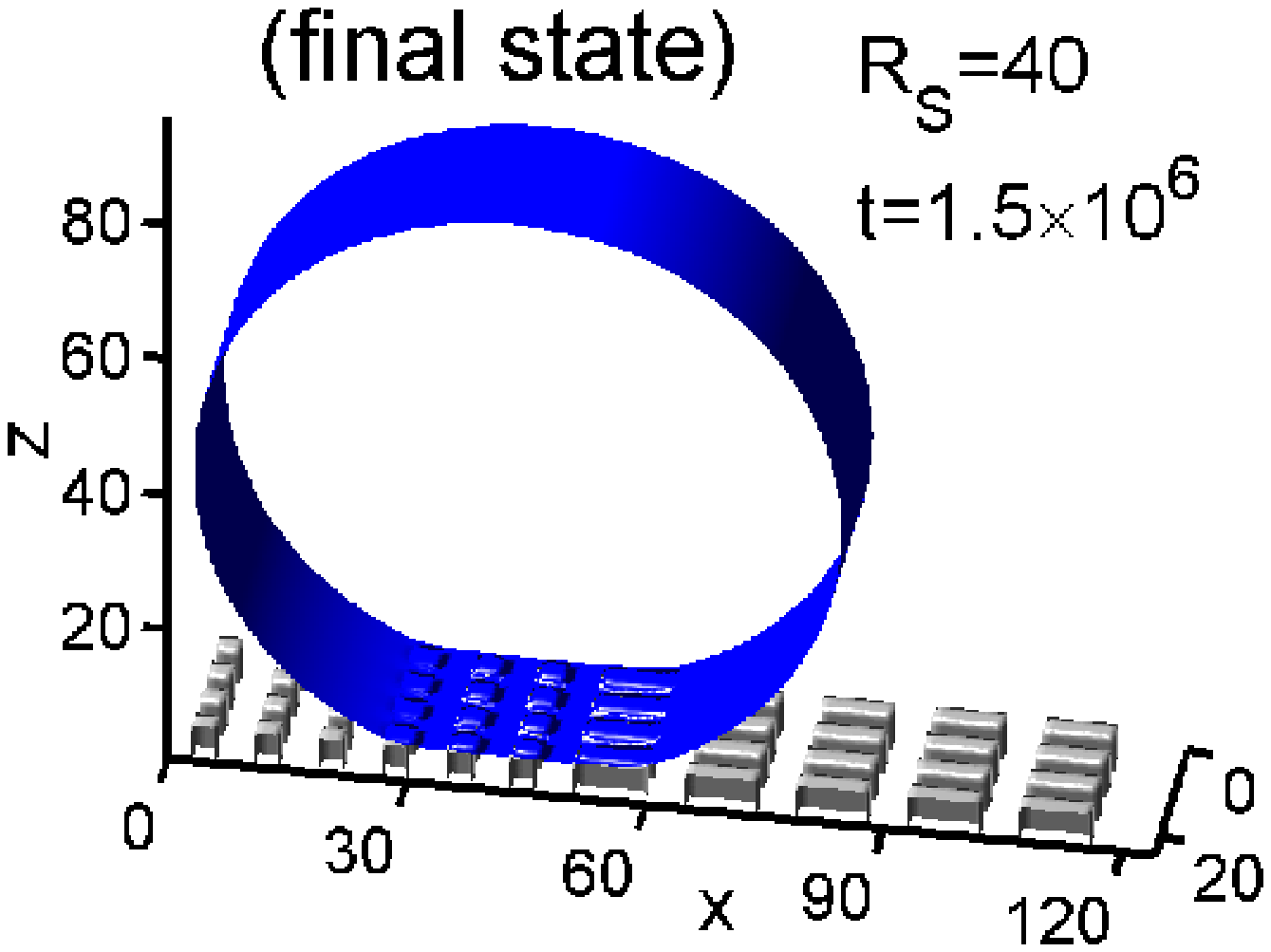}
\includegraphics[width=4.25 cm]{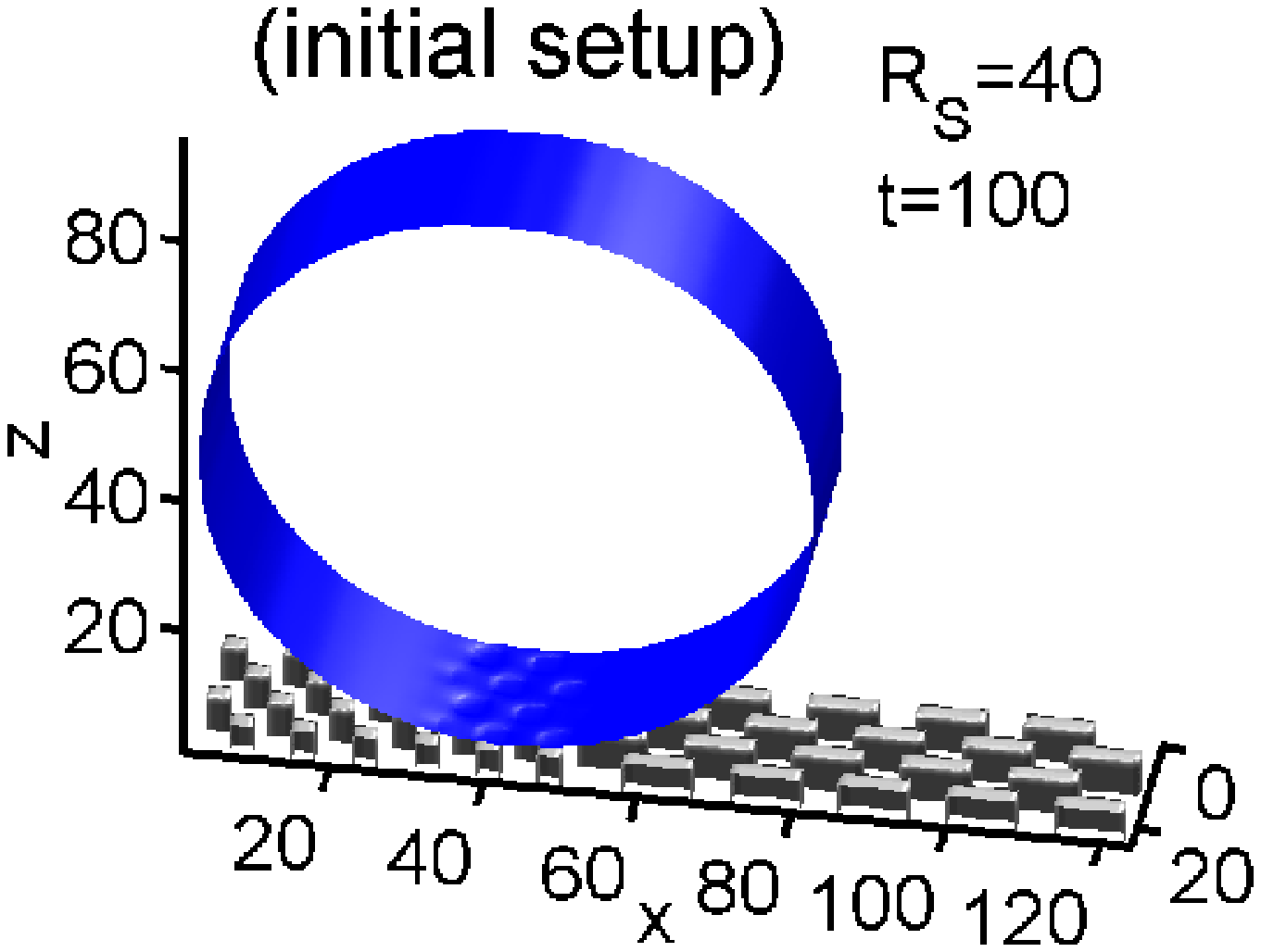}
\includegraphics[width=4.25 cm]{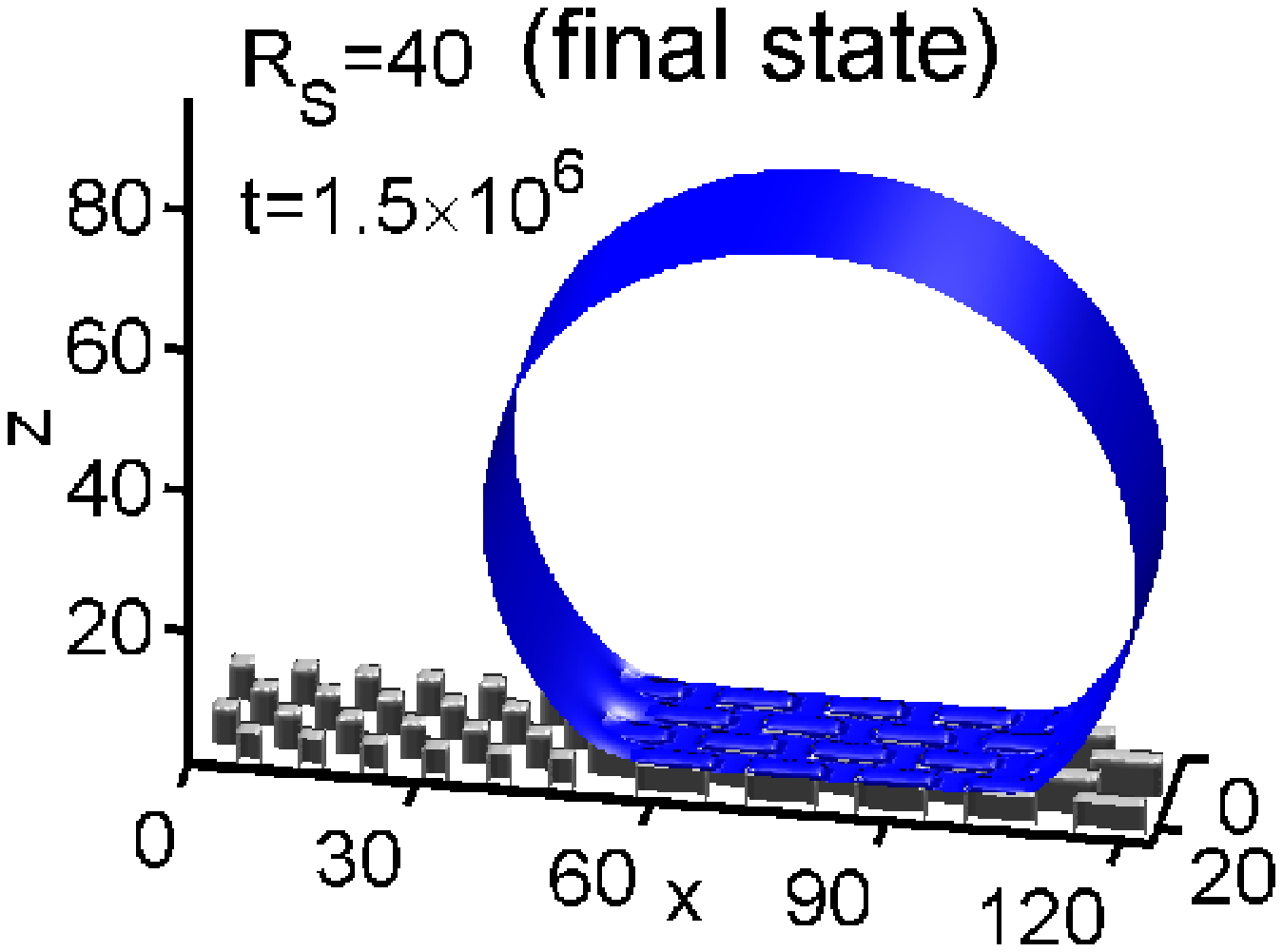}
\caption{The same set of simulations as in \figref{fig:sph_drop_snapshots} but for the case of cylindrical droplets.}
\label{fig:cyl_drop_snapshots}
 \end{figure}

In order to study the effect of droplet shape on the above phenomenon, we also performed a set of simulations using a cylindrical droplet instead of a sphere. The results of these simulations, shown in \figref{fig:cyl_drop_snapshots}, are in line with the case of spherical droplets.

\begin{figure}
 \includegraphics[width=4.2 cm]{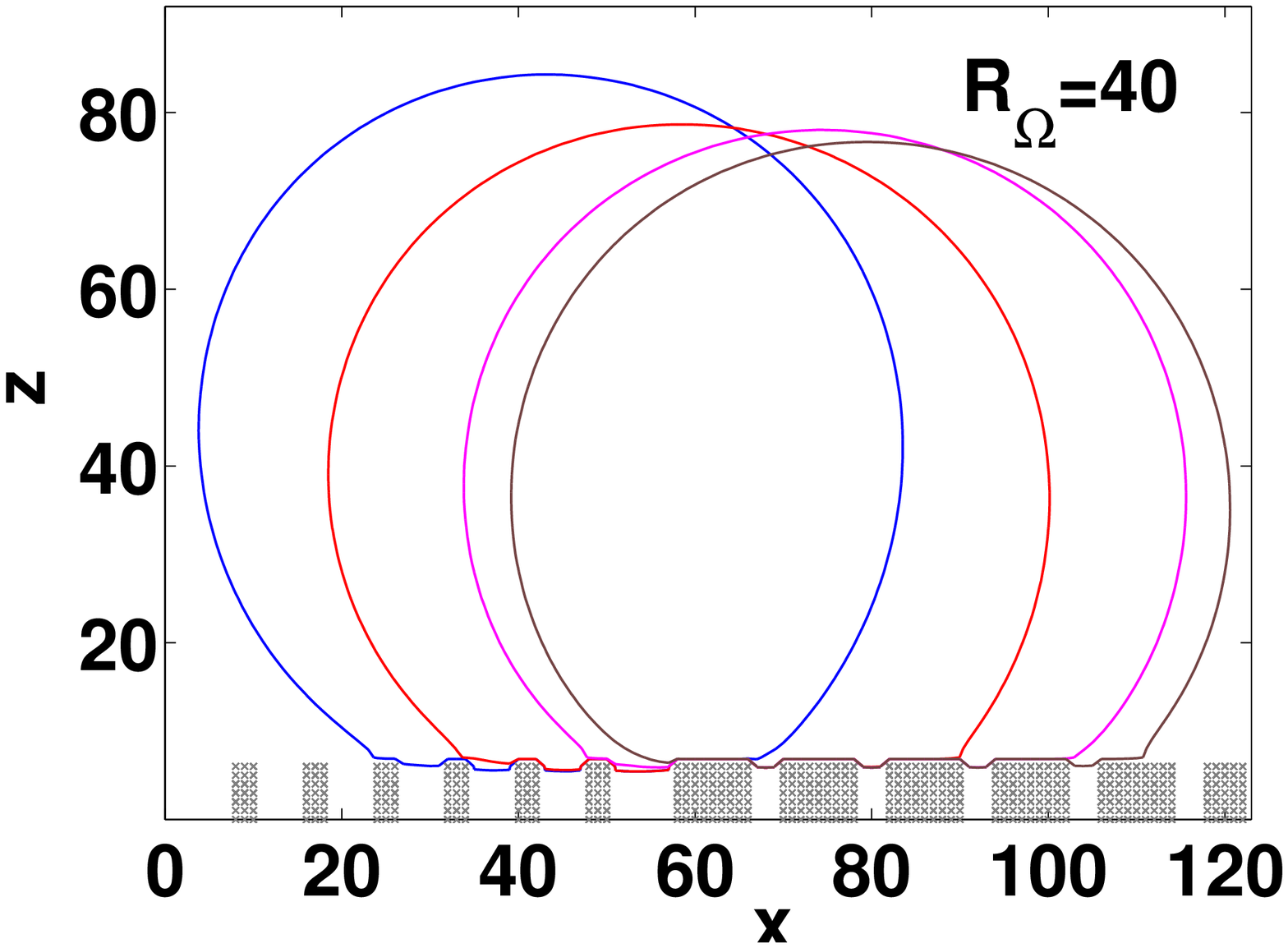}
 \includegraphics[width=4.2 cm]{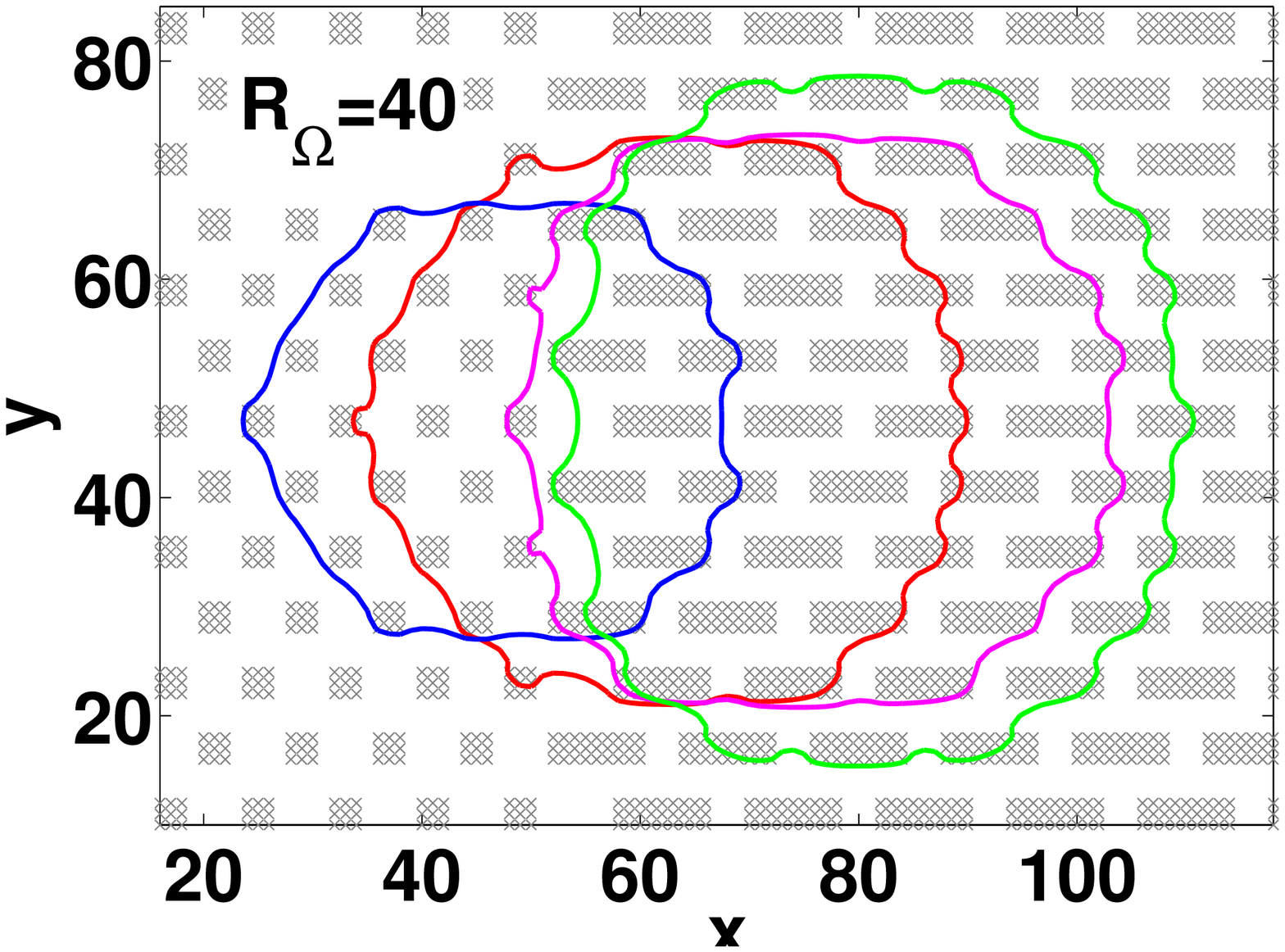}
\caption{The $xz$-cross section of the liquid-vapor interface (left) and the corresponding footprint (right) of a spherical droplet
on a step gradient substrate. In the both panels, the time increases from left to right: $t=5\times10^{4}$, $6\times10^{5}$,
$1.2\times10^{6}$ and $2\times10^{6}$.}
\label{fig05}
\end{figure}

For further considerations, we use the substrate type B. The left panel of \figref{fig05} depicts the $xz$-cross section through the center of mass of a spherical drop ($R_{\Omega}=40$) at different times during its motion over the step gradient zone (the pillar densities on the left and right halves of the substrate are fixed to $\phiLeft=0.187$ and $\phiRight=0.375$). The interested reader can see the motion of droplet in the supplementary movie. The corresponding footprint of the droplet (three phase contact line) is shown in the right panel of \figref{fig05}.

The footprint of the droplet reflects the geometry (shape and arrangement) of the posts. A trend towards increasing contact area is also observed in accordance with the lower effective contact angle in the right region (higher pillar density). A closer look at the footprints in \figref{fig05} (right panel) shows how the chess board-like arrangement of the posts allows the droplet to find the neighbor posts in the gradient direction.

\begin{figure}
\begin{center}
\includegraphics[ width=7 cm]{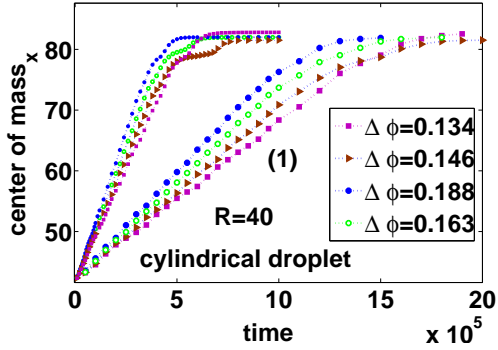}
\end{center}
\caption{The $x$-component of the center of mass position versus time for a cylindrical droplet using, $\phiLeft=0.187$ and $\phiRight=0.321$, $0.333$, $0.35$ and $0.375$. The two group of curves belong to two different surface tensions of $\sigma_0=5.4\times 10^{-4}$(LB units) (right; also labeled as (1) for further reference) and $4\sigma_0$ (left).
}
\label{fig:cyl_cm_motion}
\end{figure}

Next we create substrate patterns of type B with various values of $\Delta \phi$ by keeping $\phiLeft$ unchanged and varying $\phiRight$. Results on the dynamics of a cylindrical drop on such texture gradient substrates are shown in \figref{fig:cyl_cm_motion}. A survey of the center of mass position versus time in \figref{fig:cyl_cm_motion} reveals that the droplet motion is first linear in time until it reaches a constant value. The plateau corresponds to the case, where the droplet has completely left the region of lower pillar density. Since no driving force exists in this state, the droplet velocity vanishes due to viscous dissipation.

\begin{figure}
\begin{center}
\includegraphics[width=4.25 cm]{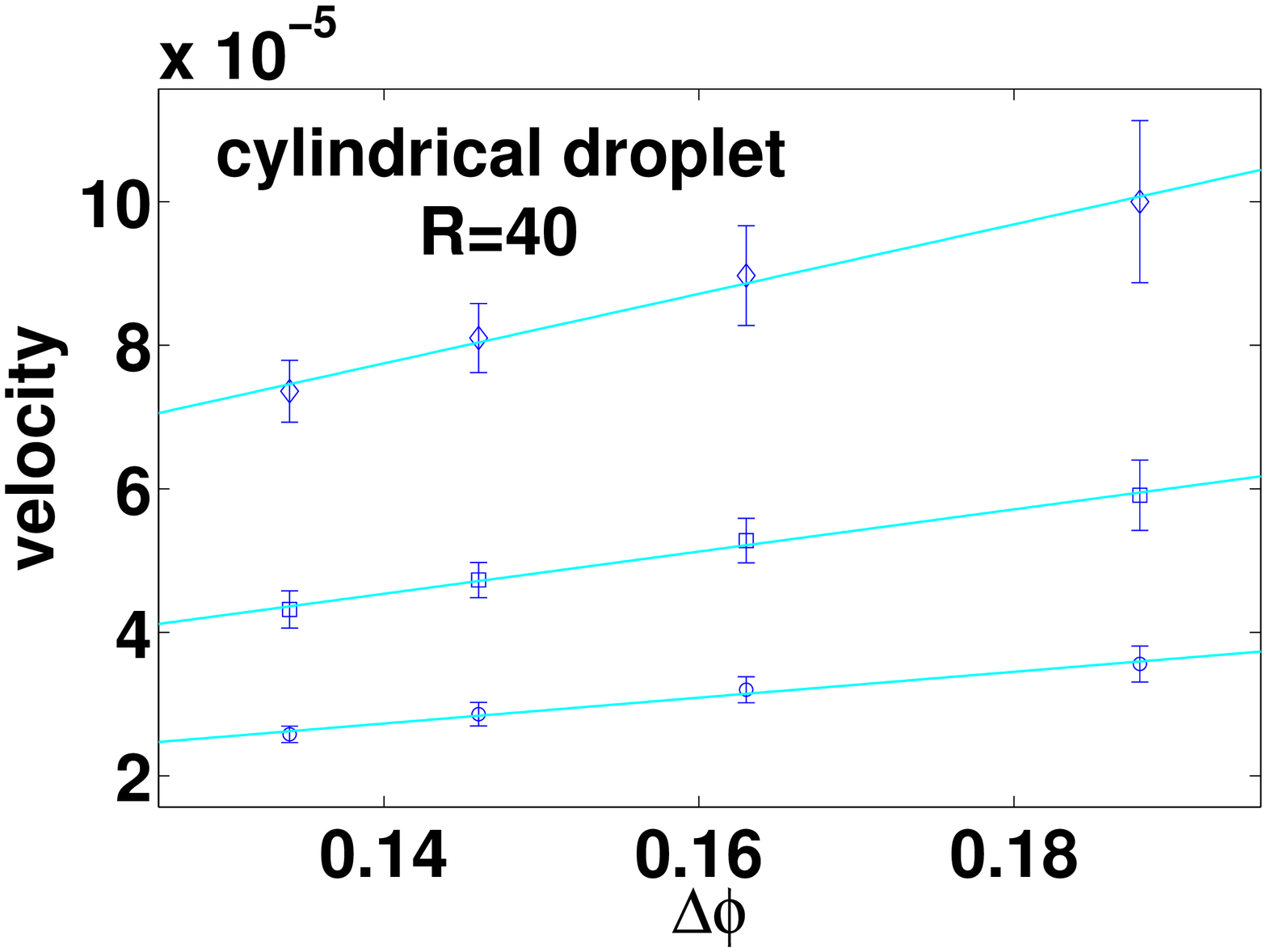}
\includegraphics[width=4.25 cm]{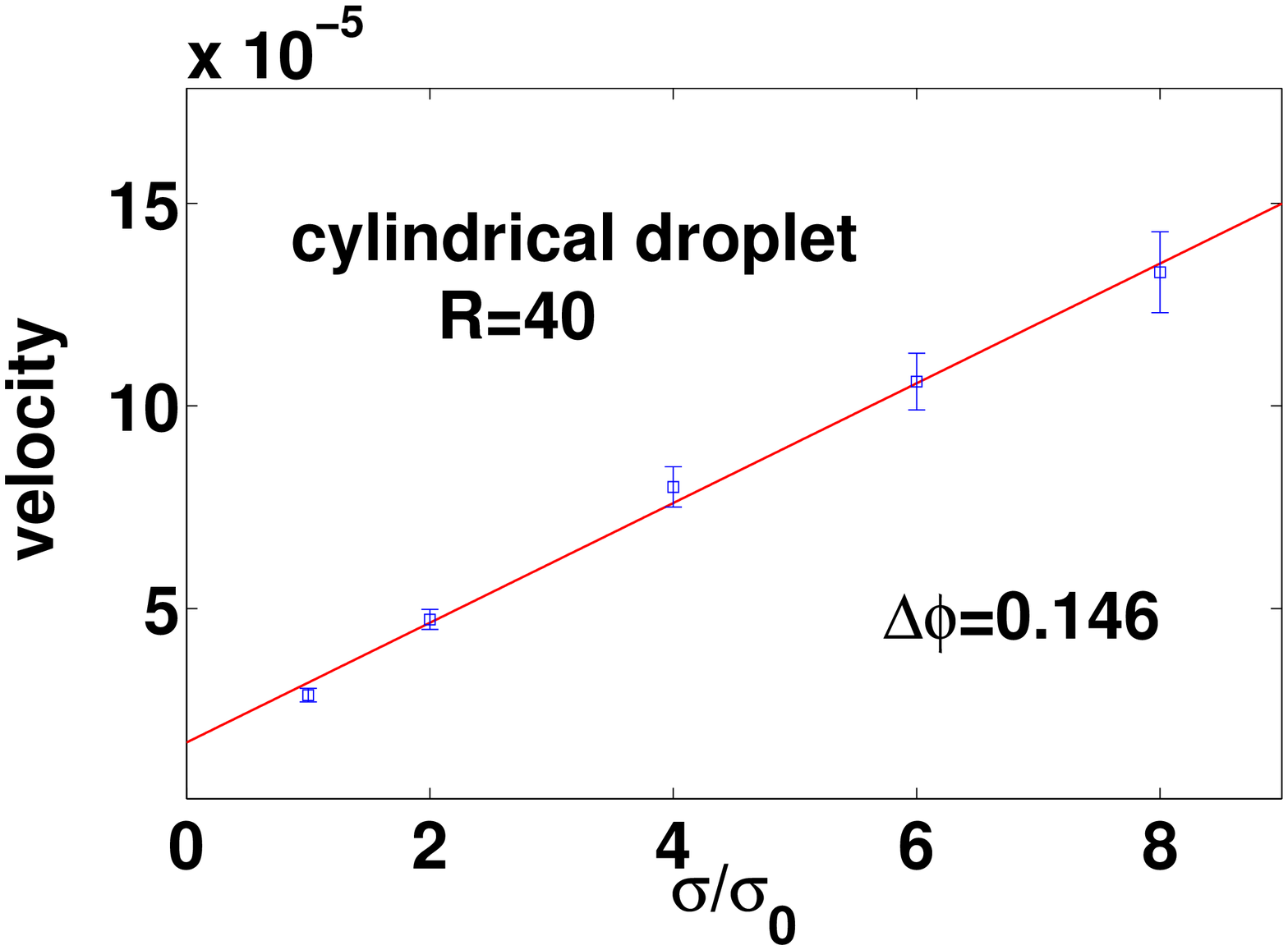}
\end{center}
\caption{Left: Droplet's center of mass velocity versus the difference in pillar density, $\Delta \phi = \phiRight-\phiLeft$, extracted from the linear part of the center of mass motion (see e.g.\ \figref{fig:cyl_cm_motion}). Results for three different liquid-vapor surface tensions are depicted. From top to bottom: $4\sigma_0$, $2\sigma_0$ and $\sigma_0$, where $\sigma_0=5.4\times 10^{-4}$ (LB units). In all cases, a linear variation is seen in accordance with the simple model, \equref{eq:velocity}.
Right: A further test of \equref{eq:velocity}, where the dependence of droplet velocity on surface tension is shown for a fixed $\Delta \phi$.}
\label{fig:velocity}
\end{figure}

Using the linear part of the data shown in \figref{fig:cyl_cm_motion}, we define an average velocity for the motion of droplet's center of mass upon the action of texture gradient forces. Importantly, \figref{fig:cyl_cm_motion} reveals the strong effect of the surface tension on droplet dynamics. Both absolute values of droplet velocity for a given $\Delta \phi$ as well as the slope of the data significantly depend upon $\sigma$.

In order to rationalize these observations, we provide a simple model based on scaling arguments. Noting that the flow we consider is in the viscous regime, we neglect inertial terms in the Navier-Stokes equation and write for the steady state $0= -\nabla p  + \eta \Delta u$, where $u$ is the fluid velocity, $p$ is the hydrostatic pressure and $\eta$ the viscosity. The velocity $u$ varies only over a distance of the order of the droplet radius, hence $\Delta u \sim u/R^2$. On the other hand, $\nabla p \sim - d\pL(\thetaC)/R =  (\sigma/R^2) dR/R$, assuming that the driving force originates from the Laplace pressure variation (over a length of the order of $R$) within the droplet. For the case of a cylindrical droplet of unit length, the condition of constant droplet volume, $\Omega=R^2 [\thetaC- \sin(2\thetaC)/2]$, \equref{eq:Cassie} and some algebra lead to $dR/R \sim (\pi-\thetaC) d \phi \sim \sqrt{\phi} d\phi$ (the relation $\pi-\thetaC\sim \sqrt{\phi}$ follows from \equref{eq:Cassie} assuming $\thetaC$ close to $\pi$ \cite{Reyssat}). Putting all together, and after a change of notation $d\phi \equiv \Delta \phi =\phiRight-\phiLeft$, we arrive at $\eta u/R^2 \sim (\sigma/R^2) \sqrt{\phi} \Delta \phi$. Hence,
\begin{equation}
u \sim  \frac{\sigma}{\eta} \sqrt{\phi} \Delta \phi.
\label{eq:velocity}
\end{equation}

Interestingly, despite different mechanisms at work, both \equref{eq:velocity} and Eq.\ (5) in \cite{Reyssat} predict a linear dependence of droplet velocity on $\Delta \phi$. In \cite{Reyssat}, the lateral velocity is estimated from roughness gradient induced asymmetry of dewetting of a droplet, flattened  due to impact. The situation we consider is different. There is no impact and hence a related flattening is absent in the present case. Furthermore, the dynamics we study is in the viscous regime whereas the high impact velocity in \cite{Reyssat} supports the relevance of inertia. These differences show up in different predictions regarding the dependence of the droplet velocity on surface tension, fluid viscosity and density. While Eq.\ (5) in \cite{Reyssat} predicts a dependence on the square root of $\sigma$, \equref{eq:velocity} suggests that, in our case, a linear dependence on $\sigma$ should be expected.

We therefore examine \equref{eq:velocity} not only with regard to the relation between droplet velocity $u$ and difference in roughness density $\Delta \phi$ (left panel in \figref{fig:velocity}) but also check how $u$ changes upon a variation of the surface tension $\sigma$ for a fixed $\Delta\phi$. Results of this latter test are depicted in the right panel of \figref{fig:velocity}, confirming the expected linear dependence of $u$ on $\sigma$. It is noteworthy that $\sigma$ in the right panel of \figref{fig:velocity} varies roughly by a factor of 10 so that a square root dependence can definitely be ruled out.

\begin{figure}
\begin{center}
\includegraphics[width=4.25 cm]{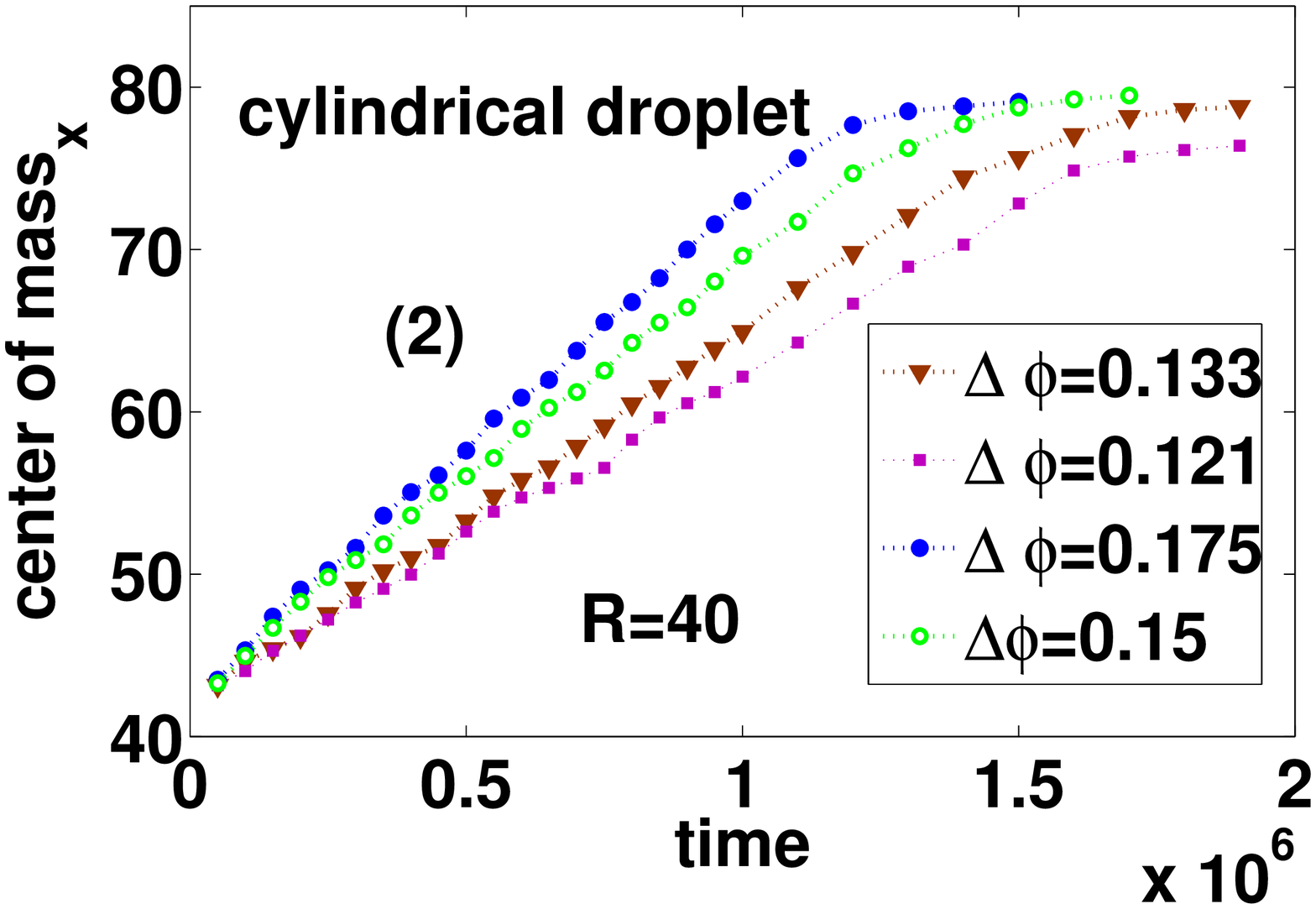}
\includegraphics[width=4.25 cm]{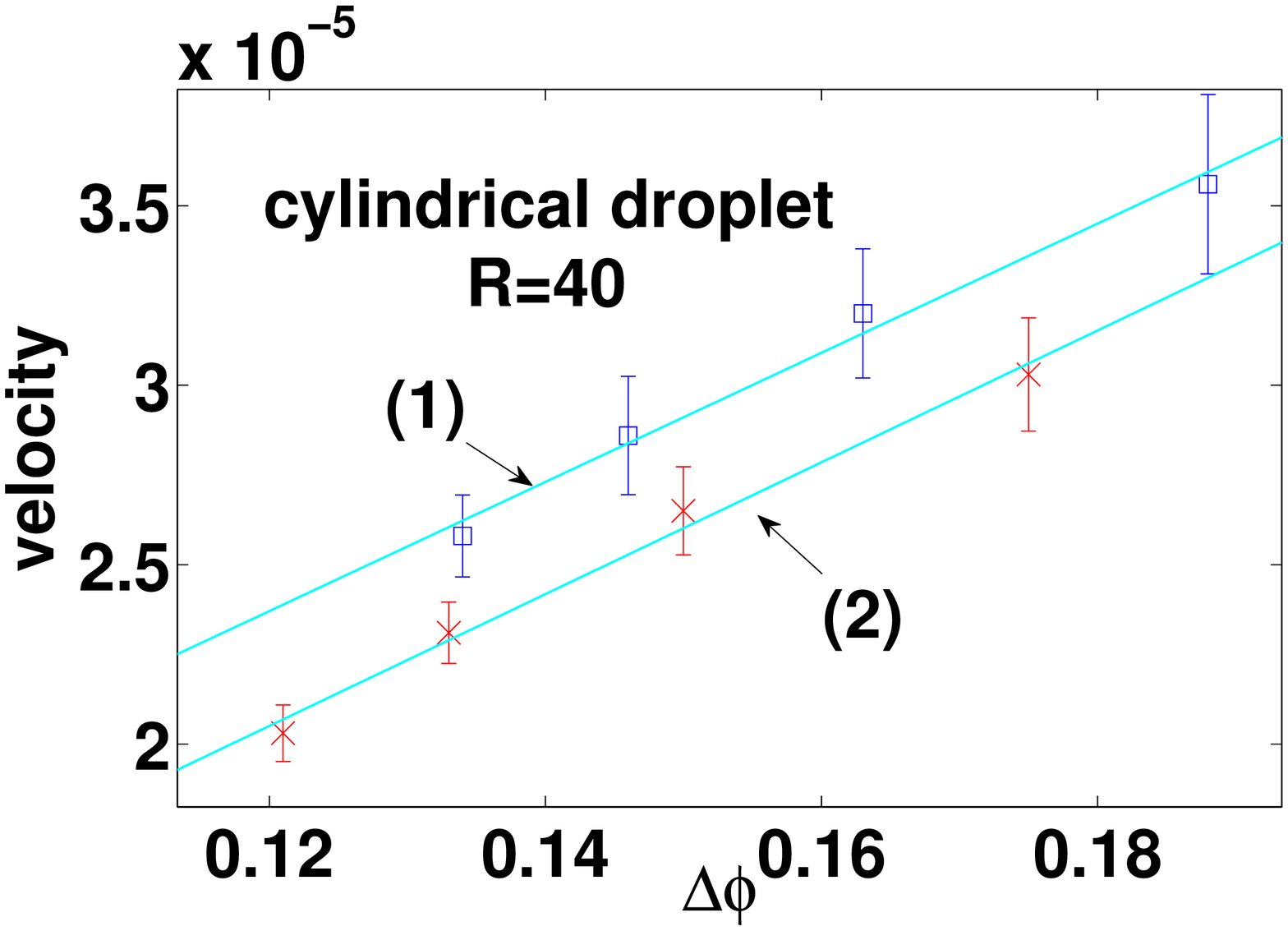}
\end{center}
\caption{Left: The $x$-component of the center of mass position versus time for a cylindrical droplet using, $\phiLeft=0.2$ and $\phiRight=0.321$, $0.333$, $0.35$ and $0.375$ [$\sigma=\sigma_0=5.4\times 10^{-4}$ (LB units)]. Right: Droplet's center of mass velocity extracted from the linear part of the data shown in the left panel and in \figref{fig:cyl_cm_motion} (labeled as (1)).}
\label{fig:velocityB}
\end{figure}

It is worth emphasizing that the above discussed linear relation between droplet velocity $u$ and difference in pillar density $\Delta \phi$ is expected to hold as long as pinning forces are weaker than texture gradient induced driving force. In this case, the specific details of pillar arrangement seem to modify the prefactors entering the scaling relation, \equref{eq:velocity}, but not the predicted linear law. \Figref{fig:velocityB} is devoted to this aspect. In this figure, the dependence of $u$ on $\Delta \phi$ is compared for two slightly different ways of realizing $\Delta\phi$: In (1) $\phiLeft=0.187$ while $\phiLeft=0.2$ in (2). All other aspects/parameters are identical. As a consequence, since the list of investigated $\phiRight$ is exactly the same, slightly higher values of $\Delta \phi$ are realized in (1) as compared to (2). If the details of pillar arrangement were unimportant, all the velocity data obtained from these two series of simulations should lie on the same line. As shown in \figref{fig:velocityB}, this is obviously not the case. Rather, the linear relation between $u$ and $\Delta\phi$ seems to hold independently for each studied case.

\begin{figure}
\includegraphics[ width=4.2 cm]{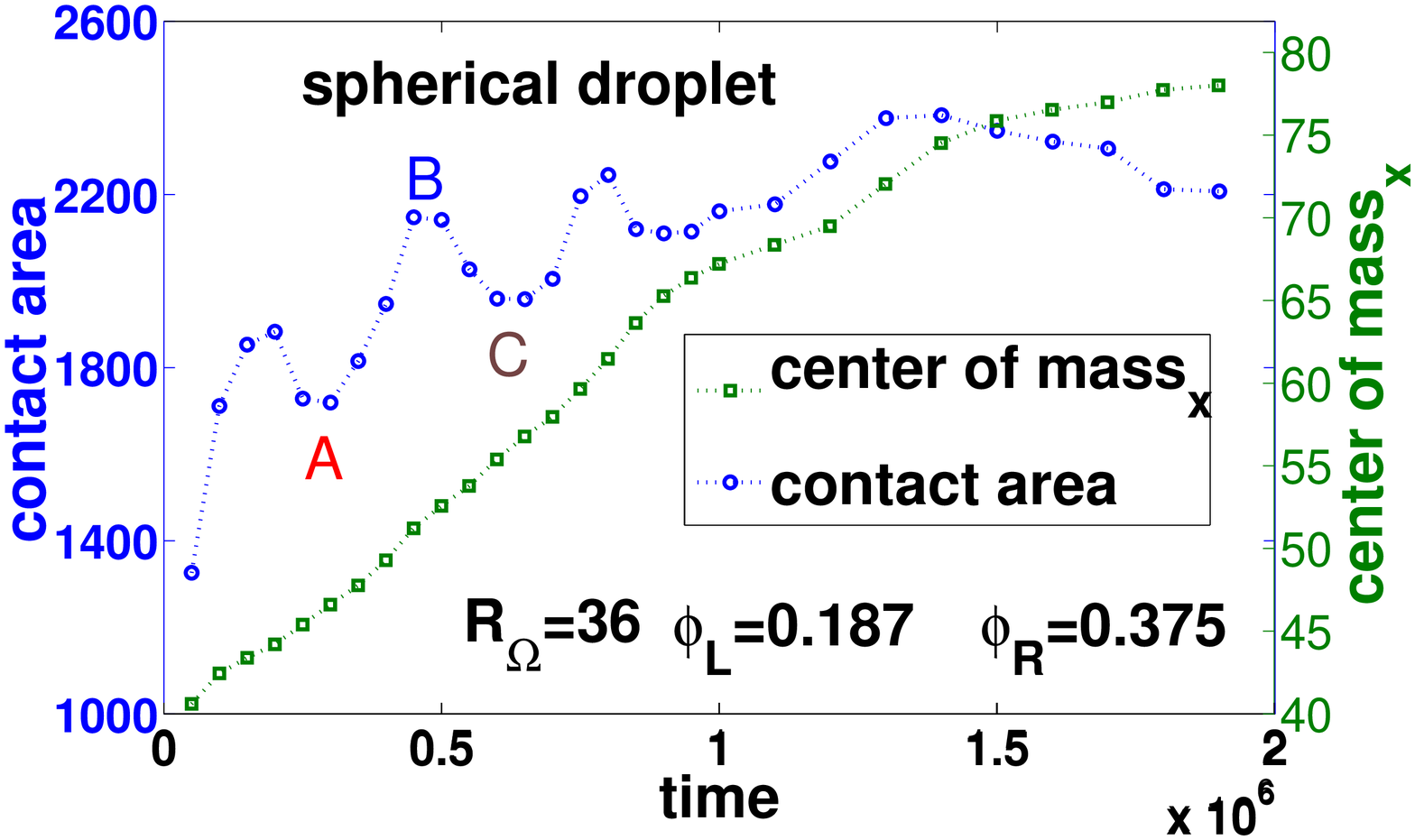}
 \includegraphics[width=4.2 cm]{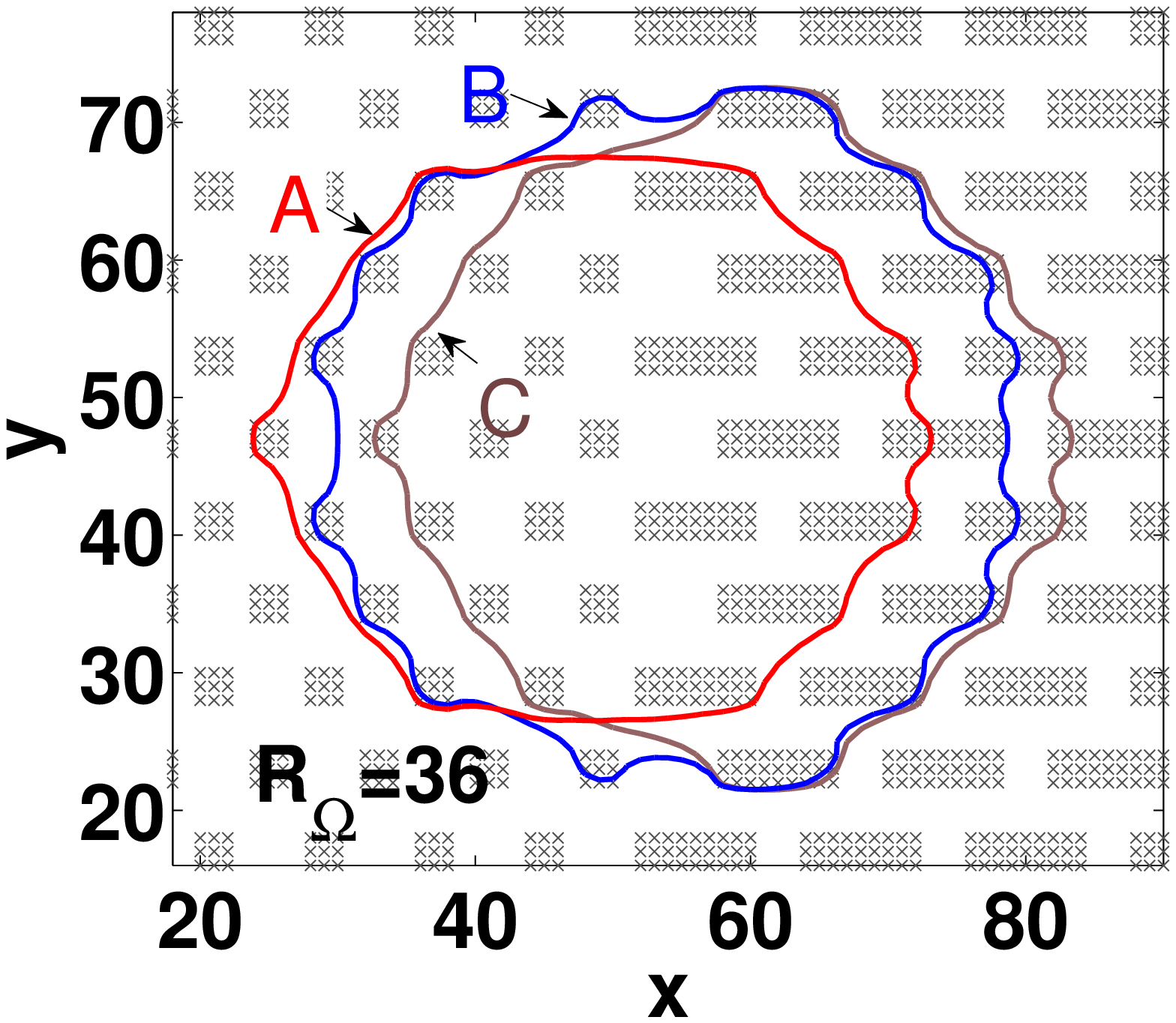}
\caption{Left: The contact area and the $x$-component of droplet's center of mass position versus time. Right: Footprint of the droplet on the step gradient substrate for times {$t_{\mathrm{A}}$, $t_{\mathrm{B}}$ and $t_{\mathrm{C}}$} corresponding to the extrema of the contact area as labeled in the left panel.}
 \label{fig06}
\end{figure}

Next we examine how the droplet's contact area changes with time as the droplet moves on the gradient zone. We determine this quantity by simply counting the number of grid points beneath the droplet. The time evolution of the contact area is compared to that of the center of mass position in \figref{fig06} (left panel) for a spherical droplet of radius $R_{\Omega}=36$. In contrast to the center of mass position, which increases monotonously with time, the area beneath the droplet exhibits irregularities and oscillations.

We presume that these irregularities and oscillations are closely related to the dynamics of three phase contact line. This idea is confirmed by the plot in the right panel of \figref{fig06}, where droplet's footprint is shown for times $t_{\mathrm{A}}$, $t_{\mathrm{B}}$ and $t_{\mathrm{C}}$ corresponding to the three extrema in the contact area labeled by A, B and C. As seen from this plot, the increase of the droplet's base area between times $t_{\mathrm{A}}$ and  $t_{\mathrm{B}}$ is accompanied by a significant motion of the three phase contact line on the right side of the footprint while it remains essentially pinned to the pillars on the left side (with the exception of depinning from the left most pillar). The state B is, however, energetically unfavorable due to a stretched shape of the contact line. The transition from B to C reduces this asymmetry, thereby leading to a smaller contact area at $t_{\mathrm{C}}$. We emphasize here that such local events are not included in the Cassie-Baxter picture, \equref{eq:Cassie}. 

\section{conclusion}
We use a two-phase lattice Boltzmann model to study the dynamic behavior of suspended droplets on patterned hydrophobic substrates with a step-wise change in pillar density. We show that the specific arrangement of pillars may play a significant role for the dynamics of the droplet on such substrates. In particular, varying the pillar arrangement while keeping the gradient of pillar density unchanged (\figref{fig:substrates}), we show that both full transport over the gradient zone as well as complete arrest between the two regions of different pillar density may occur (\figsref{fig:sph_drop_snapshots}{fig:cyl_drop_snapshots}).

The relation between the droplet motion and the gradient of pillar density is investigated, revealing a linear dependence for the range of parameters studied (\figsref{fig:cyl_cm_motion}{fig:velocity}). A simple model is provided based on the balance between the viscous dissipation and the driving force, the latter assumed as the gradient of the internal droplet (Laplace) pressure (\equref{eq:velocity}). The model not only reproduces the observed dependence on the pillar density gradient but also predicts a linear dependence of the steady state droplet velocity on the surface tension. This prediction is in line with results of lattice Boltzmann simulations, where the surface tension is varied by roughly a factor of 10 (\figref{fig:velocity}).

Moreover, comparing droplet dynamics for two slightly different ways of realizing the gradient of texture, it is shown that the gradient in pillar density does not uniquely determine the droplet velocity. Rather, the way this gradient is implemented also matters to some extent (\figref{fig:velocityB}).


A detailed survey of the contact line dynamics is also provided revealing interesting pinning and depinning events leading to small amplitude oscillations of the droplet's contact area during its motion over the gradient zone (\figref{fig06}).

\section{Acknowledgments}
We thank David Qu\'er\'e for sending us a version of his recent manuscript on gradient of texture and Alexandre Dupuis for
providing us a version of his LB code. N.M. gratefully acknowledges the grant provided by the Deutsche Forschungsgemeinschaft (DFG) under the number Va 205/3-3. ICAMS gratefully acknowledges funding from ThyssenKrupp AG, Bayer MaterialScience AG, Salzgitter Mannesmann Forschung GmbH, Robert Bosch GmbH, Benteler Stahl/Rohr GmbH, Bayer Technology Services GmbH and the state of North-Rhine Westphalia as well as the European Commission in the framework of the European Regional Development Fund (ERDF).


\begin{thebibliography}{99}
\bibitem{Quere}          de Gennes P. G., Brochard-Wyart F.  and Qu\'{e}r\'{e} D., Capillarity and Wetting Phenomena, Springer  (2004).
\bibitem{Dorrer}        Dorrer C. and  R\"{u}he J., Soft Matter \textbf{5},  (2009) 51.
\bibitem{QuereAnnu}     Qu\'{e}r\'{e} D., Annu. Rev. Mater. Res. \textbf{38}, (2008) 71.
\bibitem{Ajdari}        Stone H. A., Stroock A. D.  and Ajdari A., Annu. Rev. Fluid. Mech. \textbf{36}, (2004) 381.
\bibitem{Prakash}       Prakash M.  and Gershenfeld N., Science \textbf{315}, (2007) 832.
\bibitem{Rabbe}         Varnik F.  and Raabe D., Modelling Simul. Mater. Sci. Eng. \textbf{14}, (2006) 857.
\bibitem{Young}         Young T., Trans. Roy. Soc. \textbf{95}, (1805) 65.
\bibitem{Wenzel}        Wenzel R. N., Ind. Eng. Chem. \textbf{28}, (1936) 988.
\bibitem{Cassie}        Cassie A. B. D.  and Baxter S., Trans. Faraday Soc. \textbf{40}, (1944) 546.
\bibitem{Gao2007a}      Gao L.  and McCarthy T. J., Langmuir \textbf{23}, (2007) 3762.
\bibitem{Reyssat1}      Reyssat M., Yeomans J. M.  and  Qu\'{e}r\'{e} D., EPL \textbf{81}, (2008) 26006.
\bibitem{Jopp}          Jopp J., Gr\"{u}ll  H.  and Yerushalmi-Rozen  R., Langmuir \textbf{20}, (2004) 10015.
\bibitem{Markus}        Gross M., Varnik F.  and Rabbe D., EPL, (2009)  accepted.
\bibitem{Li}            Li X. M., Reinhoudt D.  and Grego-Calama M., Chem. Soc. Rev \textbf{36}, (2007) 1350.
\bibitem{Oner}          \"{O}ner D.  and  McCarthy T. J., Langmuir \textbf{16}, (2000) 7777.
\bibitem{Joanny}        Joanny J. F. and de Gennes P. G., J. Chem. Phys. \textbf{81}, (1984) 552.
\bibitem{Lafuma}        Lafuma A.  and Qu\'{e}r\'{e} D., Nature Materials \textbf{2}, (2003) 457.
\bibitem{Yang}          Yang J. T., Chen J. C., Huang K. J.  and  Yeh J. A., J. Microelectromech. Syst. \textbf{15}, (2006) 697.
\bibitem{Zhu}           Zhu L., Feng  Y., Ye X.  and Zhou Z., Sensors and Actuators A \textbf{130-131}, (2006) 595.
\bibitem{Shastry}       Shastry A., Case  M. J.  and  B\"{o}hringer K. F., Langmuir \textbf{22}, (2006) 6161.
\bibitem{Reyssat}       Reyssat M., Pardo F.  and Qu\'{e}r\'{e} D., EPL \textbf{87}, (2009) 36003.
\bibitem{Fang}          Fang G., Li  W., Wang X. and Qiao G., Langmuir \textbf{ 24}, (2008) 11651.
\bibitem{Swift}         Swift M. R., Osborn W. R.  and Yeomans J. M., Phys. Rev. Lett.\textbf{ 75}, (1995) 830.
\bibitem{Holdych}       Holdych D. J., Rovas D., Georgiadis J. G.  and Buckius R. O., Int. J. Mod. Phys. C \textbf{9}, (1998) 1393.
\bibitem{Briant}        Briant A. J., Wagner A. J.  and Yeomans J. M., Phys. Rev. E \textbf{69}, (2004) 031602.
\bibitem{Kusumaatmaja}  Kusumaatmaja H., Blow M. L., Dupuis A.  and  Yeomans J. M., EPL \textbf{81}, (2008) 36003.
\bibitem{Dupuis}        Dupuis A.  and Yeomans J. M., Europhys. Lett. \textbf{75}, (2006) 105.
\bibitem{Yeomans}       Yeomans J. M. and Kusumaatmaja H., Bull. Pol. Ac.: Tech. \textbf{55}, (2007) 203.
\bibitem{Kusumaatmaja2} Kusumaatmaja H., L\'eopold\`es J., Dupuis A.  and  Yeomans J. M., Europhys. Lett. \textbf{73}, (2006) 740.
\bibitem{VarnikF}       Name Varnik F., Truman P., Wu B., Uhlmann P., Raabe D.  and  Stamm M., Phys. Fluids \textbf{20}, (2008) 072104.


\bibitem{Kusumaatmaja3}  Kusumaatmaja H. and Yeomans J. M., Langmuir \textbf{23}, (2007) 6019.


\end{thebibliography}
\end{document}